\documentclass[a4paper,11pt]{article}
\pdfoutput=1 

\usepackage{jcappub} 

\usepackage{amssymb}
\usepackage{times}
\usepackage{graphicx}
\usepackage{epstopdf}
\usepackage{subfigure}
\usepackage[T1]{fontenc}
\usepackage{aecompl} 
\usepackage{multirow}
\usepackage{pdflscape}
\usepackage{rotating}

\usepackage{anyfontsize}
\usepackage{hyperref}

\newcommand{\beq}{\begin{equation}}
\newcommand{\eeq}{\end{equation}}

\def\gs{\mathrel{\lower0.6ex\hbox{$\buildrel {\textstyle >}\over{\scriptstyle \sim}$}}}
\def\ls{\mathrel{\lower0.6ex\hbox{$\buildrel {\textstyle <}\over{\scriptstyle \sim}$}}}
\newcommand{\simgt}{\lower.5ex\hbox{$\; \buildrel > \over \sim \;$}}
\newcommand{\simlt}{\lower.5ex\hbox{$\; \buildrel < \over \sim \;$}}

\newcommand{\aap}{A\&A}

\newcommand{\apj}{ApJ}

\newcommand{\apjs}{ApJS}
\newcommand{\aj}{AJ}

\newcommand{\mnras}{MNRAS}

\newcommand{\ssr}{Space Science Reviews}

\newcommand{\mdash}{-}
\newcommand{\gt}{$\gs$}

\title{\boldmath Comparison of weak lensing by NFW and Einasto halos and systematic errors}

\author[a,b,1]{Mauro Sereno,\note{Corresponding author.}}
\author[b,c]{Cosimo Fedeli,}
\author[a,b,c]{Lauro Moscardini}

\affiliation[a]{Dipartimento di Fisica e Astronomia, Alma Mater Studiorum -- Universit\`a di Bologna, viale Berti Pichat 6/2, 40127 Bologna, Italia}
\affiliation[b]{INAF, Osservatorio Astronomico di Bologna, via Ranzani 1, 40127 Bologna, Italia}
\affiliation[c]{INFN, Sezione di Bologna, viale Berti Pichat 6/2, 40127 Bologna, Italia}

\emailAdd{mauro.sereno@unibo.it}

\abstract{Recent $N$-body simulations have shown that Einasto radial profiles provide the most accurate description of dark matter halos. Predictions based on the traditional NFW functional form may fail to describe the structural properties of cosmic objects at the percent level required by precision cosmology. We computed the systematic errors expected for weak lensing analyses of clusters of galaxies if one wrongly models the lens density profile. Even though the NFW fits of observed tangential shear profiles can be excellent, viral masses and concentrations of very massive halos ($\gs 10^{15}M_\odot/h$) can be over- and underestimated by $\sim 10$ per cent, respectively. Misfitting effects also steepen the observed mass-concentration relation, as observed in multi-wavelength observations of galaxy groups and clusters. Based on shear analyses, Einasto and NFW halos can be set apart either with deep observations of exceptionally massive structures ($\gs 2\times10^{15}M_\odot/h$) or by stacking the shear profiles of thousands of group-sized lenses ($\gs 10^{14}M_\odot/h$).
}


\begin{document}
\maketitle
\flushbottom

\section{Introduction}

The hierarchical cold dark matter paradigm with a cosmological constant ($\Lambda$CDM) is highly successful in describing the properties of the Universe and the evolution and formation of structure therein. Dark matter halos are traditionally modeled as Navarro-Frenk-White (NFW) density profiles \citep{nav+al97,ji+su02}. As with many other ingredients of the $\Lambda$CDM cosmology, the main reasons behind assuming NFW halos are that they result from $N$-body simulations and they work relatively well in modeling observable properties of galaxies and clusters.

High precision astrophysics nowadays requires that a level of accuracy of $\sim$ 1--2 per cent on the measurement of density profiles of dark matter halos is achieved. Recent investigations (see for instance \cite{sal08,du+ma14,kly+al14,men+al14} and references therein) have shown that CDM haloes are described best by the Einasto density profile, which is given by $d\ln \rho/d \ln r \propto r^\alpha$ with a constant value of $\alpha$ \citep{ein65}. NFW and Einasto halos differ mainly at very small radial distances from the center of very massive structures. These regions are difficult to probe observationally, and as a consequence it is generally assumed that NFW profiles can work as well as Einasto models.

Problems, however, might arise when highly non-linear processes are involved. Deviations of the NFW functional form from $N$-body results are generally small for halos that are not very massive, but can be significant for massive clusters at high redshift \citep{kly+al14}. Halos that are well fitted by NFW models in 3D may be not NFW-like in projection \citep{men+al14}. This is caused by the halo triaxiality and by the effects of substructures and additional matter along the line of sight. Furthermore, observable quantities are generally connected to the gravitational potential and its derivates rather than the mass distribution. A reasonable characterization of the 3D density profile may then fail to reproduce all the observable features of a halo. A description of the halo in terms of the potential could be more sensible \citep{tch+al15}.

Well known systematic effects can bias the measurement of mass and concentration in weak lensing analyses and the proper comparison to theoretical predictions. The presence of a dominant central galaxy has to be accounted for in studies of the cluster core. Baryonic effects can influence the dark matter halo profiles \citep{fed12}. Irregular clusters may exhibit substructures and deviations from a simple parametric profile \citep{men+al10}. Halo concentration can be over-estimated in halos elongated along the line of sight \citep{lim+al13,ser+al13}. Cluster member galaxies can be erroneously treated as background source galaxies due to wrongly assigned photometric redshifts \citep{kol+al15}. Staked analyses can be affected as well by scatter in the observable mass proxy and by scatter in concentration at fixed mass. Miscentering effect on small scales and the 2-halo term on large scales have to be considered too \citep{ser+al15_bias}. Finally, selection effects can plague the analysis of not statistical samples of clusters \citep{men+al14,ser+al15_cM}.

The low level of statistical uncertainties expected in ongoing and upcoming large area surveys demands for a proper treatment of all sources of systematic errors. In this paper we quantify the systematic errors made by modeling the observable weak lensing properties of Einasto halos by means of NFW profiles. The NFW models fail mainly in irregular clusters out of equilibrium. For these systems our results will provide a conservative estimate of the minimum error. In our analysis we hence assumed that dark matter halos are in fact Einasto-like, and that we have been wrongly modeling them with the NFW parameterization.

Weak lensing analyses provide reliable mass measurements \citep{wtg_III_14,ume+al14,se+et15_comalit_I,ser15_comalit_III}. Notwithstanding the long history of gravitational lensing, lensing by Einasto haloes is still in its infancy \citep[ and references therein]{ret+al12a,ret+al12b}. The main reason is that the properties of Einasto lenses cannot be expressed in terms of simple functions. \cite{mam+al10} computed some approximated interpolating functions. \cite{ret+al12a} and \cite{ret+al12b} expressed the convergence and the shear in terms of Fox $H$ functions. The complementing analysis of the lensing properties of the Sersic model can be found in \cite{car04}. 

On the observational side, \cite{man+al08} verified that Einasto and NFW modelings gave consistent results within errors in the analysis of stacked lensing from galaxies and clusters in the Sloan Digital Sky Survey (SDSS). \cite{ume+al14} described the stacked tangential shear signal of 20 massive CLASH clusters \citep[Cluster Lensing And Supernova survey with Hubble,][]{pos+al12}. They found that the shape parameter, mass and concentration of the Einasto model were statistically consistent with the NFW-equivalent parameters. The Einasto profile together with other dark matter models was also tested with stacked gravitational lensing of high-mass clusters in \cite{ber+al13}.

We adopted as a reference background model the 2013 Planck cosmology \citep{planck_2013_XVI}, with $\Omega_\mathrm{M}=1- \Omega_{\Lambda}=0.3175$, Hubble constant $H_0=100h~\mathrm{km~s}^{-1}\mathrm{Mpc}^{-1}$ with $h=0.671$, and $\sigma_8=0.8344$ for the amplitude of the linear matter power spectrum, which we computed  as outlined in \cite{ei+hu98}.

\section{Halo models}

The Einasto profile  \citep{ein65} provides an excellent fit to dark matter simulated halos,
\begin{equation}
\label{ein1}
	\rho_\mathrm{Ein}=\rho_{-2} \exp \left\{ -\frac{2}{\alpha}\left[ \left(\frac{r}{r_{-2}}\right)^\alpha -1\right] \right\},
\end{equation}
where $\alpha$ is the shape parameter, $r_{-2}$ is the radius where the logarithmic slope of the density profile is $-2$, and $\rho_{-2}$ is the density at $r_{-2}$. The NFW density profile follows the functional form \citep{nav+al97},
\begin{equation}
\label{nfw1}
	\rho_\mathrm{NFW}=\frac{\rho_\mathrm{s}}{(r/r_\mathrm{s})(1+r/r_\mathrm{s})^2},
\end{equation}
where $r_\mathrm{s}$($=r_{-2}$) is the scale radius and $\rho_\mathrm{s}$($=4\rho_{-2}$) is the scale density. 

A halo can be generally described in terms of its mass and concentration. We adopted $M_{200}$, the mass inside the sphere of radius $r_{200}$, wherein the mean density is 200 times the critical density at the halo redshift. The concentration is defined as $c_{200} \equiv r_{200}/ r_\mathrm{-2}$. This definition may contrast with the common-sense notion that the more concentrated object is the one that has the denser central region and the less dense outer halo. In fact, the mass inside the scale radius of the Einasto halo also depends on the shape parameter. An alternative definition is connected to the ratio of the maximum circular velocity to the virial velocity, $V_\mathrm{max}/V_{200}$ \citep{pra+al12,kly+al14}. 

The Einasto and NFW profiles are quantitatively different at small and large radii. On scales of interest for weak gravitational lensing ($0.1\ls r/r_{200}\ls1.0$) the NFW profile can only mimic an Einasto profile with $\alpha\sim0.2$ \citep{du+ma14}. Einasto halos have three parameters, one more than NFW. This additional degree of freedom is not the reason behind their better performances \citep{kly+al14}. In fact, the logarithmic slope is related to the peak height, defined as $\nu(M_{200},z) \equiv \delta_\mathrm{c}/\sigma(M_{200},z)$, and it increases with mass \citep{gao+al08}. In the following, we fixed the shape parameter to $\nu$ as \citep{kly+al14},
\begin{equation}
\label{eq_ein2}
	\alpha = 0.115+0.0165\nu^2 .
\end{equation}

\section{Einasto versus NFW lensing}

\begin{figure}[tbp]
\centering
\begin{tabular}{c}
\includegraphics[width=.6\textwidth]{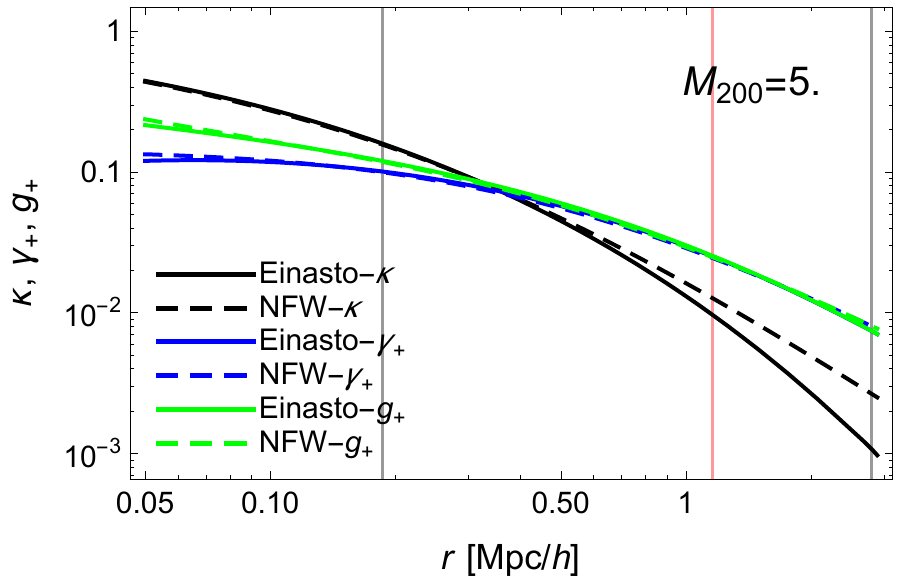} \\
\includegraphics[width=.6\textwidth]{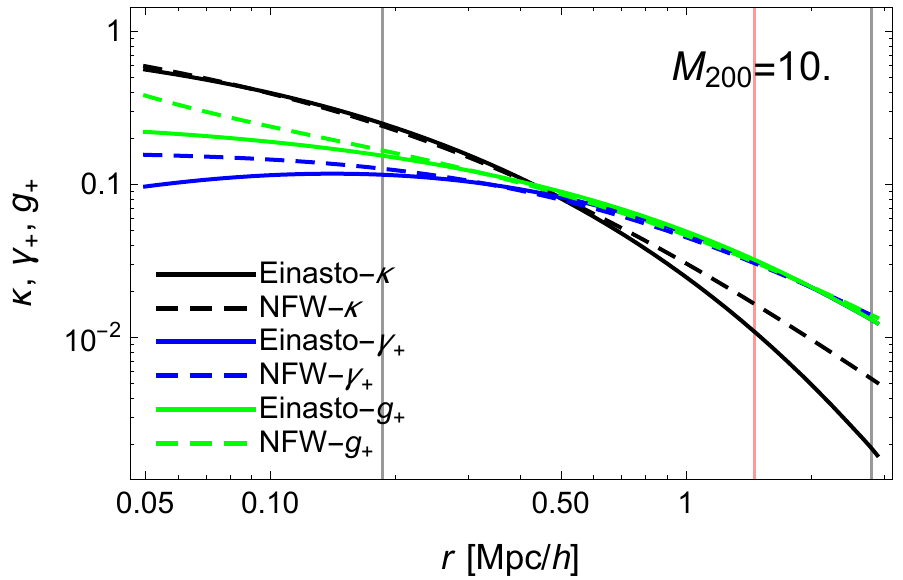} \\
\includegraphics[width=.6\textwidth]{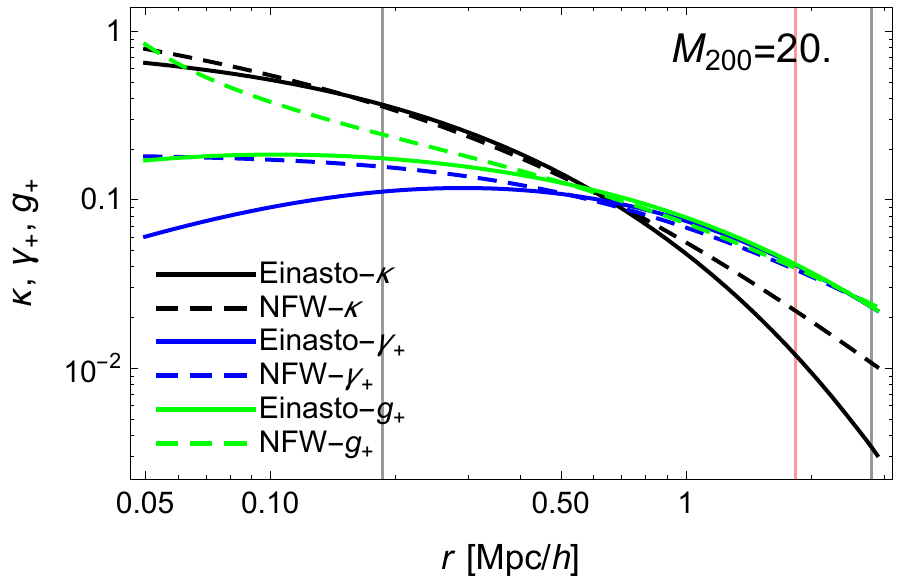} \\
\end{tabular}
\caption{Profiles of convergence (black), tangential shear (blue), and reduced tangential shear (green). The full and dashed lines are for the Einasto and the NFW profiles, respectively. The vertical gray lines bracket the radial range between 1' and 15'. The vertical red line marks $r_{200}$. The lens is at $z_\mathrm{d}=0.3$, the sources at $z_\mathrm{s}=1.0$. {\it Top panel}: $M_{200}=5 \times 10^{14}M_\odot/h$ and $c_{200}=4.0$. {\it Middle panel}: $M_{200}=10 \times 10^{14}M_\odot/h$ and $c_{200}=3.7$. {\it Bottom panel}: $M_{200}=20 \times 10^{14}M_\odot/h$ and $c_{200}=3.4$.}
\label{fig_shear_M200}
\end{figure}

Let us now compare the lensing properties of NFW halos to Einasto lenses with the same mass and concentration, see Figure~\ref{fig_shear_M200}. Concentrations were associated to masses following the $c-M$ relation for Einasto halos derived in \cite{du+ma14}. For the normalization, we considered the lens at redshift $z_\mathrm{d}=0.3$ and the background sources at $z_\mathrm{s}=1.0$. 

Differences in convergence, i.e. the lensing renormalized projected surface density, are more pronounced at large radii whereas differences in the reduced tangential shear profiles $g_+$ show up at small radii.

For $M_{200}=5 \times 10^{14}M_\odot/h$, the NFW convergence is similar to the true (Einasto) convergence up to $r\ls 5~\mathrm{Mpc}/h$ and then overestimates it. On the other hand, the reduced shear profile is underestimated by $\ls5$ per cent between $0.2\ls r\ls 1~\mathrm{Mpc}/h$. At very small ($r\sim 0.05~\mathrm{Mpc}/h$) or very large ($r\sim 3~\mathrm{Mpc}/h$) radii, $g_+$ is overestimated by $\sim$ 10 per cent. 

The lensing properties of NFW and Einasto lenses are very similar to each other in a wide radial range for a lens with $M_{200}=10^{15}M_\odot/h$. The NFW reduced shear reproduces the real profile within $\sim 5$ per cent in the range $0.2\ls r\ls 3~\mathrm{Mpc}/h$. Differences are most notable in very massive halos. The reduced shear of lenses with $M_{200}=2 \times 10^{15}M_\odot/h$ is overestimated by $\gs 10$ per cent in the central regions ($r\sim 0.3~\mathrm{Mpc}/h$) and underestimated by $\ls10$ per cent at $r\sim 1~\mathrm{Mpc}/h$. On the other hand, the convergence is well reproduced up to $r\sim 1~\mathrm{Mpc}/h$.

Whereas shear profiles mostly differ at small radii, the convergences are unlike at large radii. The combined analysis of galaxy tangential distortion and the complementary lensing magnification \citep{ume+al14} can be then effective in distinguishing Einasto from NFW lenses.

\section{Method}
\label{sec_meth}

In order to estimate the systematic errors experienced by wrongly modeling the observed lensing properties of a cluster, we fitted the properties of the Einasto halos with NFW profiles and compared the results of the regression to the true parameters.

We characterized the simulated Einasto halo of mass $M_{200}$ with a concentration given by the median $c-M$ relation in the Planck cosmology \citep{du+ma14}. The shape parameter was related to the mass with Eq.~(\ref{eq_ein2}). As a main test we considered lensing observations of the reduced tangential shear profile $g_+$. The $\chi^2$ function in this case is 
\beq
\label{eq_fit2D}
\chi_\mathrm{2D}^2 =\sum_i \left[ \frac{g_\mathrm{Ein,+}(\theta_i)-g_\mathrm{NFW,+}(\theta_i; M_\mathrm{2D,200},c_\mathrm{2D,200})}{\delta_{+}(\theta_i)}\right]^2,
\eeq
where $g_{+}$  is measured (with statistical uncertainty $\delta_+$) in circular annuli at angular position $\theta_i$. Expressions for the shear induced by a NFW halo can be found in \cite{wr+br00}. Lensing by Einasto deflectors is discussed in \cite{ret+al12a,ret+al12b}.

We considered observational weak lensing conditions obtainable with deep lensing ground-based programs  such as CLASH or the Sloan Giant Arcs Survey \citep[SGAS,][]{hen+al08,ogu+al12}. We calculated the shear in 10 discrete radial bins spanning the range $[\theta_\mathrm{2D,min}=1', \theta_\mathrm{2D,max}=15']$ with a constant logarithmic angular spacing. The redshift of clusters was selected to be $z_\mathrm{d}=0.3$. The observational uncertainty $\delta_{+}$ was computed by assuming a background source population with surface density of $n_\mathrm{g}=20$ galaxies per square arcminute at $z_\mathrm{s}=1.0$ and a dispersion in galaxy intrinsic ellipticities of $\sigma_\epsilon=0.3$. We added the cosmic noise due to uncorrelated large scale structure projected along the line of sight following \cite{hoe03}. Non-linear evolution was computed as outlined in \cite{smi+al03}. We associated errors to each shear measurement but we did not scatter the shear estimates. Thus the expected $\chi^2$ for the true model is zero.

As a second test, we considered the 3D fit of the density profile. We minimized the function:
\begin{equation}
\label{eq_fit3D}
	\chi^2_\mathrm{3D}=\sum_i \left[  \ln \rho_\mathrm{Ein}(r_i) - \ln \rho_\mathrm{NFW}(r_i; M_\mathrm{3D, 200},c_\mathrm{3D, 200})  \right]^2,
\end{equation}
where the density $\rho$ was measured at 50 radial positions from the cluster center spanning the range $[r_\mathrm{3D,min}=0.01r_{200}, r_\mathrm{3D,max}=r_{200}]$ with a constant logarithmic radial spacing. The 3D profile is not an observable like the shear profile. Fitting procedures like Eq.~(\ref{eq_fit3D}) are however routinely employed in studies of $N$-body simulated samples to study halo concentrations \citep{du+ma14,men+al14}.

\section{Results}

\subsection{Mass and concentration estimates}

\begin{figure}
\centering
\begin{tabular}{c}
\includegraphics[width=.7\textwidth]{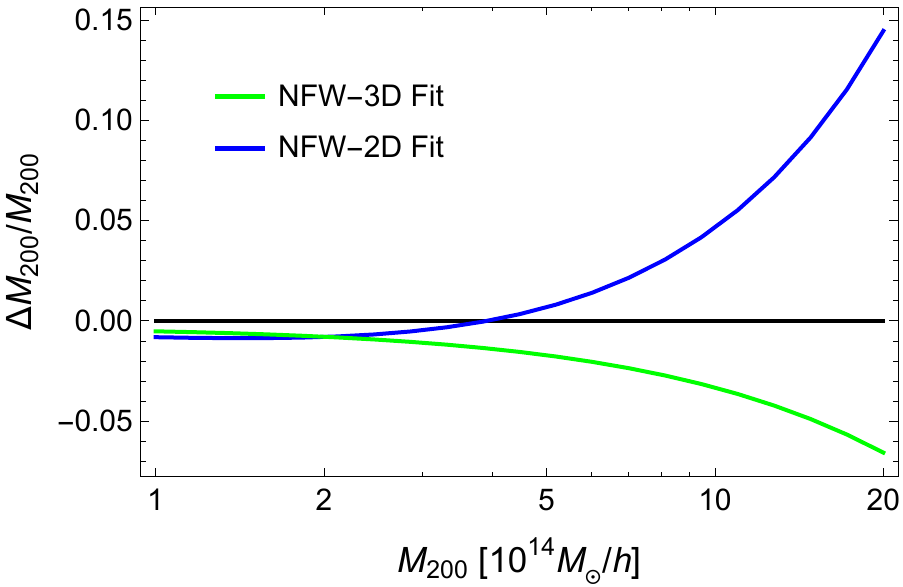} \\
\includegraphics[width=.7\textwidth]{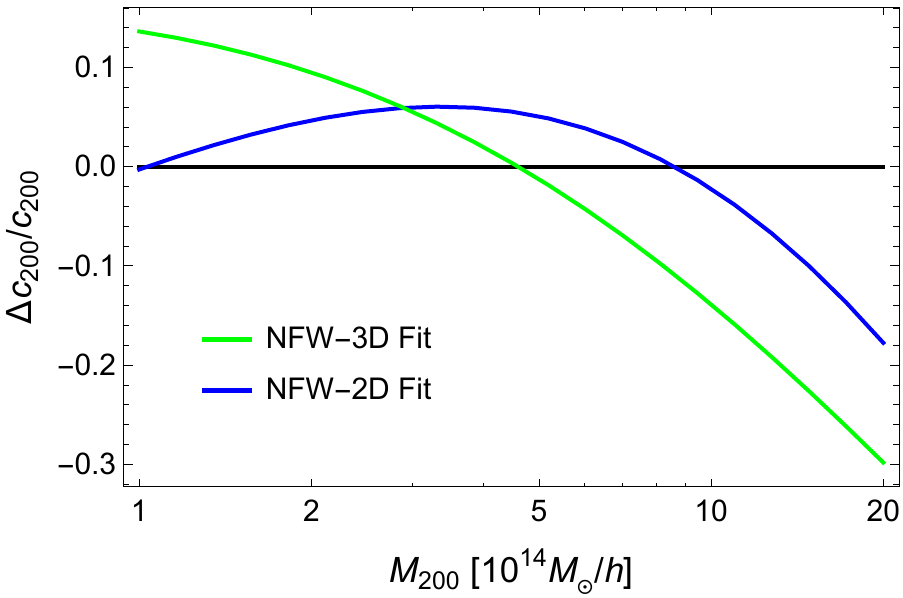} \\
\end{tabular}
\caption{Relative systematic errors as a function of mass made when fitting Einasto halos with a NFW functional form. The blue and green lines refer to the fits of the reduced shear profile and to the fit of the 3D density profile, respectively. {\it Top panel}: Relative errors in mass. {\it Bottom panel}: Relative errors in concentration.}
\label{fig_Delta_M200}
\end{figure}

The 3D fit consistently underestimates the masses, see upper panel of Figure~\ref{fig_Delta_M200}. The bias is small (less than 2 per cent) up to $M_{200}=5 \times 10^{14}M_\odot/h$, but can be significant (up to $10$ per cent) for very massive halos ($M_{200}\sim 2 \times 10^{15}M_\odot/h$). Mass estimates from the 3D or the 2D fit are very similar for low masses ($1\ls M_{200}\ls 3 \times 10^{14}M_\odot/h$). On the other hand, the 2D fit overestimates masses larger than $M_{200}\sim 5 \times 10^{14}M_\odot/h$. 

Errors in the determination of the concentration are larger, see bottom panel of Figure~\ref{fig_Delta_M200}. The 3D fit severely overestimates $c_{200}$ by $\gs 10$ per cent for low mass halos ($M_{200}\sim 10^{14}M_\odot/h$) and significantly underestimates $-$ by nearly 30 per cent $-$ the concentrations of very massive objects ($M_{200}\sim 2 \times 10^{15}M_\odot/h$). 

Errors expected in the 2D fit of the concentration are usually smaller but they can be still significant. The error is negligible at $M_{200}\sim 10^{14}$ or  $\sim 10^{15}M_\odot/h$. The maximum overestimate ($\sim 7$ per cent) is for $M_{200}\sim 3-4 \times 10^{14}M_\odot/h$, whereas the concentration of clusters with $M_{200}\gs 5 \times 10^{14}M_\odot/h$ is underestimated. Errors in masses and concentrations are most significant at the high mass tail of the cluster distribution ($M_{200}\gs 10^{15}M_\odot/h$). Very massive clusters are the preferential targets of dedicated weak lensing programs. Nearly half the clusters in either the CLASH or the WtG \citep[Weighing the Giants,][]{wtg_I_14} programs have a mass in excess of $M_{200}\sim 10^{15}M_\odot/h$.

\subsection{Model identification}

\begin{figure}
\centering
\includegraphics[width=.7\textwidth]{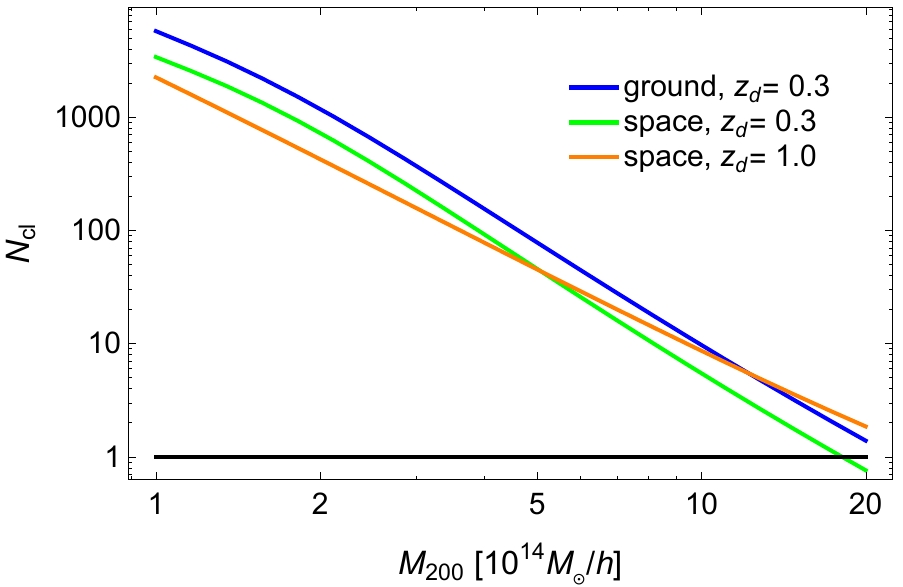}
\caption{Number $N_\mathrm{cl}$ of clusters of mass $M_{200}$ that have to be stacked to distinguish Einasto from NFW shear profiles for different surveys and different lens redshifts $z_\mathrm{d}$.}
\label{fig_Ncl_M200}
\end{figure}

Einasto and NWF models are difficult to distinguish based on weak lensing properties only. In fact, NFW models provide excellent fits to Einasto shear profiles although with biased masses and concentrations. 


We can compute the minimum number of clusters of given mass that have to be stacked to reach positive evidence for model identification, i.e. a difference of 2 for the Bayesian information criterion \citep{lid04}, see Figure~\ref{fig_Ncl_M200}. 

We considered ground- (see Section~\ref{sec_meth}) or space-based surveys. A Euclid-like survey\footnote{\url{http://sci.esa.int/euclid/}} can reach $\sigma_\epsilon=0.25$ and an effective number density of $n_\mathrm{g}\sim 30$ galaxies per square arcminute. For high-$z$ lenses at $z_\mathrm{d}=1.0$, we considered 10 angular bins between 0.5' and 5' and a median source redshift of $z_\mathrm{s}\sim 1.5$. The inner radius follows the choice made in the analysis of space-based weak lensing profiles in \cite{jee+al11}, whereas the outer radius accounts for the extended coverage obtainable in large surveys. 


Even for deep observations, the peculiar lensing signature can be identified in single clusters only if the halo is exceptionally massive. Halo identification is difficult at high redshifts too. Compared to the intermediate redshift lenses, the number of high-$z$ lenses to stack is smaller at low masses and it is larger for very massive haloes.

Our estimate of the number of clusters to be stacked is conservative. The impact of cosmic noise, which is a major source of uncertainty at large radii and high redshifts \citep{gru+al15}, can be reduced by modeling the uncorrelated large-scale structure with data readily available in large photometric and spectroscopic surveys \citep{hoe+al11b}. Furthermore, stacking increases the signal-to-noise ratio in the very inner regions ($\theta <1'$), which we did not consider in the fitting of low redshift clusters, where differences between Einasto and NFW are more prominent. An optimized radial binning more refined in the inner regions can also weight more the differences.

On the other hand, these gains are counterbalanced by off-centering effects and intrinsic scatter in the observable proxy of the mass used to bin the lenses, which smooths out the signal \citep{ser+al15_bias}.

\subsection{Mass--concentration relation}

\begin{figure}
\centering
\includegraphics[width=.7\textwidth]{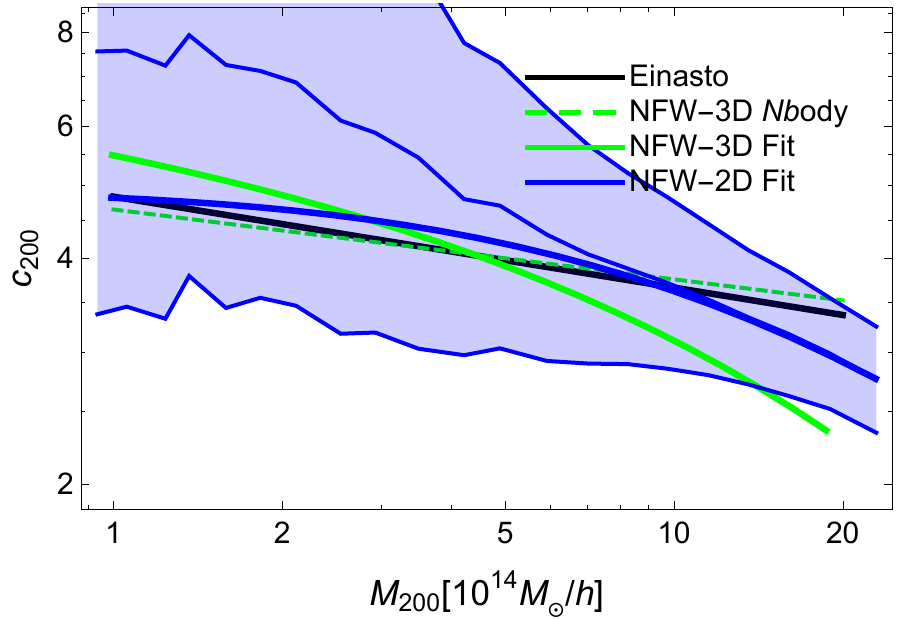} \\
\caption{The mass-concentration relation. The black (dashed green) line is the input (approximated) $c$-$M$ relation derived in \cite{du+ma14} by fitting simulated halos with Einasto (NFW) profiles. The green solid line is the $c$-$M$ relation obtained by fitting the 3D density profile of the Einasto halo with a NFW profile. The blue lines are the $c$-$M$ relations obtained by fitting the reduced shear profile of the Einasto halo with the NFW model. The thick and thin blue lines denote the best fit results (i.e. which approximates the relation estimated in absence of observational errors) and the bi-weight estimator of the marginalized posterior probability distribution. The shaded blue area includes the 1-$\sigma$ confidence region.
}
\label{fig_cM_Einasto}
\end{figure}


Erroneously analyzing Einasto shear profiles with the NFW functional form steepens the observed $c-M$ relation, see Figure~\ref{fig_cM_Einasto}. The measured relation is overconcentrated at low masses and underconcentrated at the large mass tail with respect to the input relation. Errors are within the statistical uncertainties but the effect is systematic. The bias in the slope estimation is strongly sensitive to the inner minimum radius considered in the fitting procedure. We conservatively considered $\theta_\mathrm{min}=1'$. The smaller, $\theta_\mathrm{min}$, the larger the bias. The steepening is present also for the 3D fit of the density profiles.

The presence of the steepening does not depend on the assumed effective density of background sources and the related observational uncertainty on the measured shear. Even in case of infinitely accurate and precise measurements, when the estimated concentration (thin blue line in Figure~\ref{fig_cM_Einasto}) tends to the best fit value (thick blue line), the measured $c-M$ is steeper.

The effect we measured is larger than the difference between NFW (green-dashed line in Figure~\ref{fig_cM_Einasto}) and Einasto (black-full) 3D fits to $N$-body simulated haloes \citep{du+ma14}. This is due to the details of the fitting procedure and to the fact that $N$-body simulated dark matter halos are not exactly Einasto. 


\section{Conclusions}

Profiles which excellently fit the 3D density distribution of clusters of galaxies can fail to predict the observable projected properties of the corresponding halos at the level of 1--2 per cent required by precision cosmology. We estimated the systematic errors in the measurement of mass and concentration made by fitting Einasto-like shear profiles with NFW models. Effects can be prominent for very massive halos with virial masses in excess of $M_{200}\sim 10^{15}M_\odot/h$. Misfitting effects can cause  overestimations of the mass by $\sim 10$ per cent and underestimation of the concentration by $\sim 10$ per cent.

Einasto and NFW halos cannot be yet distinguished by present stacked analyses. Accidentally, \cite{man+al08} and \cite{ume+al14} considered mass scales ($M_{200} \gs 10^{14}$ and $\sim 10^{15}M_\odot/h$, respectively) where the difference is minimal. Future Euclid-like surveys are going to provide the required accuracy.

The accurate determination of the lens profile can also benefit by combining shear and magnification analyses \citep{ume+al14}. Count-depletion measurements using flux-limited samples can constrain the convergence profile and help to break degeneracies.

Misfitting effects can help to solve the problem of observed massive lensing clusters following a steep scaling in tension with predictions from the concordance $\Lambda$CDM paradigm \citep{co+na07,ser+al15_cM}. This tension can be mostly solved  by considering the strong anti-correlation between lensing measured mass and concentration \citep{aug+al13,ser+al15_cM}, the adiabatic contraction of the halos and the presence of a dominant brightest cluster galaxy \citep{fed12,ser+al15_cM,gio+al12a,gio+al14}, and, mostly, selection effects \citep{ogu+al05,men+al14,ser+al15_cM}. The effect we considered goes in the same direction.

\acknowledgments

MS and LM acknowledge financial contributions from contracts ASI/INAF I/023/12/0, PRIN MIUR 2010-2011 'The dark Universe and the cosmic evolution of baryons: from current surveys to Euclid', and PRIN INAF 2012 'The Universe in the box: multiscale simulations of cosmic structure'. CF has received funding from the European Commission Seventh Framework Programme (FP7/2007-2013) under grant agreement No. 267251. 



\providecommand{\href}[2]{#2}\begingroup\raggedright\endgroup

\end{document}